\documentclass{article}
\usepackage[dvips]{graphics}

\begin{document}

\title{Mapping dust column density in dark clouds by using NIR scattered light : case of the Lupus 3 dark cloud}

\author{Yasushi \textsc{Nakajima}, Ryo  \textsc{Kandori}, Motohide  \textsc{Tamura} \\
{\small National Astronomical Observatory of Japan} \\
Tetsuya  \textsc{Nagata} \\ 
{\small Kyoto University}\\
Shuji  \textsc{Sato}\\
{\small Nagoya University} \\
and \\
Koji  \textsc{Sugitani} \\
{\small Nagoya City University} }

\maketitle

\begin{abstract}
We present a method of mapping dust column density in dark clouds 
by using near-infrared scattered light.
Our observations of the Lupus 3 dark cloud indicate that 
there is a well defined relation between (1) the $H-K_s$ color of an individual 
star behind the cloud, i.e., dust column density, and 
(2) the surface brightness of scattered light toward the star in each of 
the $J$, $H$, and $K_s$ bands.   
In the relation, the surface brightnesses increase at low $H-K_s$ colors, 
then saturate and decrease with increasing $H-K_s$.
Using a simple one-dimensional radiation transfer model, 
we derive empirical equations which plausibly represent the observed relationship 
between the surface brightness and the dust column density.
By using the empirical equations, we estimate dust column density of the cloud 
for any directions toward which even no background stars are seen. 
We obtain a dust column density map with a pixel scale  
of 2.3 $\times$ 2.3 arcsec$^2$ and a large dynamic range up to $A_V$ = 50 mag.  
Compared to the previous studies by Juvela et al., 
this study is the first to use color excess of the background stars 
for calibration of the empirical relationship and to apply the empirical relationship 
beyond the point where surface brightness starts to decrease with increasing column density.
\end{abstract}

\section{Introduction}

Dark clouds are seen as dark patches against bright star fields, while 
they are observed as reflection nebulae at near-infrared (NIR) wavelengths 
in the recent decade (Lehtinen \& Mattila 1996, Nakajima et al. 2003, Foster \& Goodman 2006).  
The images of NIR reflection nebulae give us an insight that darker areas with fewer background stars 
have larger column densities. In our previous paper (Nakajima et al. 2003),  we showed 
that the Lupus 3 dark cloud appears as an NIR reflection nebula and suggested 
that there is a relationship between the NIR surface brightness and dust column density toward 
the Lupus 3 dark cloud. 
It is worth investigating whether if we can quantify the relationship between the NIR surface brightness 
and column density.  

Star counting, color excess of background stars, molecular gas and thermal dust emissions have been 
used for column density mapping of dark clouds. Juvela et al. (2006) reviewed these methods intensively. 
Each of these methods has limitations. A common limitation to all these methods is that it is hard to obtain 
a high spatial resolution. 
Padoan et al. (2006) and Juvela et al. (2006) examined the use of scattered NIR surface brightness 
as a new high resolution tracer of the interstellar clouds. 
They derived an analytical formula to convert NIR surface brightness to visual extinction.  
The formula is linear at low visual extinctions and starts to saturate when visual extinction reaches 
$\sim$ 10 mag.
Juvela et al. (2006) used numerical simulations 
and radiative transfer calculations to show that the NIR surface brightness can be converted to 
column density in the range of $A_V$ $<$ 15 mag with accuracy of better than 20\%. 
Juvela et al. (2008) applied the method to a filamentary cloud in Corona Australis. 
The method using surface brightness and one using color excess of background star 
agrees below $A_V$ $\sim$ 15 mag. 

In this paper, we propose an empirical method of estimating column density of dark clouds 
up to $A_V$ $\sim$ 50 mag by using NIR surface brightness. 
We re-examine the NIR data of the Lupus 3 dark cloud in Nakajima et al. (2003) and 
obtain a plausible empirical equations which fit the observed relationship between 
the surface brightness and column density. 
Using the relationship, we obtain the dust column density map from the surface brightness.

\begin{figure}
\begin{center}
\includegraphics{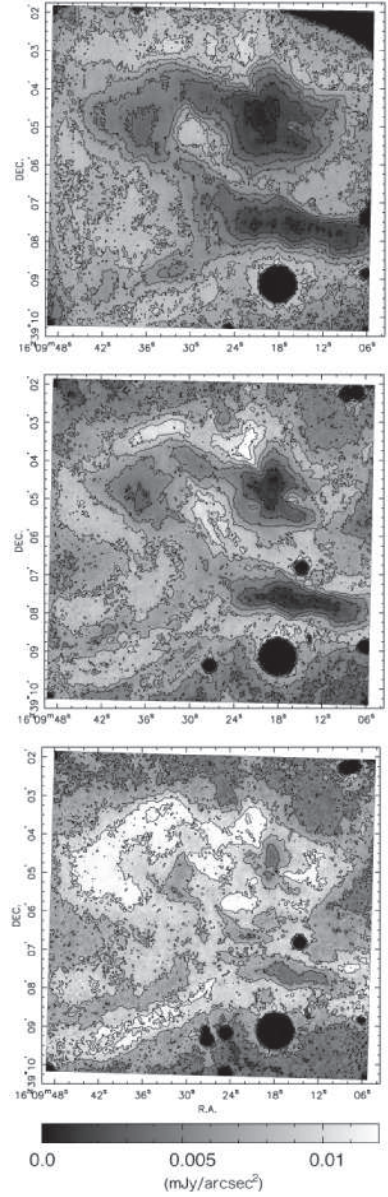}
\caption{Surface brightness of the Lupus 3 dark cloud in the $J$, $H$, and $K_s$ bands from top to bottom. 
  Saturated stars are masked with circles in black. Bad pixel clusters in the J band are also shown 
 in black. The lowest contour level denotes the 3-sigma limiting flux and the interval of contour is 
 set to the 3-sigma limiting flux value at each band. }\label{fig:fig1}
\end{center}
\end{figure}

\section{Data}

We obtained the $J$, $H$, and $K_s$ band data of the Lupus 3 dark cloud on 2001 June 4 and 7 
with the IRSF 1.4 m telescope and the NIR camera SIRIUS. The total integration time was 135 minutes 
for each band. Procedures of observations and data reduction are almost the same as described 
in Nakajima et al. (2003). The differences are as follows.   
We included data obtained on 2001 June 7 in addition to those on 2001 June 4 
to obtain a higher signal-to-noise ratio. 
We calibrated photometry by comparing stars in the observed field with 2MASS catalog stars instead of 
using standard stars. 
Color conversion factors between the 2MASS and IRSF/SIRIUS system were taken into account (Nakajima et al. 2008). 
Hence, magnitudes and colors are presented in the 2MASS photometric system. 
We selected well-measured stars by the sharpness value of ALLSTAR routine in IRAF \footnote{IRAF is 
distributed by the National Optical Astronomy Observatories, which are operated by the Association of 
Universities for Research in Astronomy, Inc., under cooperative agreement with the National Science Foundation.}. 
We rejected stars with an absolute value of sharpness larger than 0.2 as no point-like sources.  
The total number of the stars which were selected in both the $H$ and $K_s$ bands was 2910 
and that in all the three bands was 1611. 
We corrected vertical streaks due to bright stars by comparing adjacent columns. 
We calculated the surface brightness toward each pixel by taking the median value 
of 5 $\times$ 5 = 25 pixels centered at the pixel after subtraction of stars by using SUBSTAR routine in IRAF. 
The pixel area corresponds to 2.3 $\times$ 2.3 arcsec$^2$, and the error was estimated 
by standard deviation of the 25 pixels. 
The 3-sigma limiting fluxes for surface brightness were 
$F_J = 1.1$, $F_H = 1.6$, and $F_{K_s}= 2.3$ $\mu$Jy arcsec$^{-2}$.  

\begin{figure}
  \begin{center}
  \scalebox{0.7}{\includegraphics{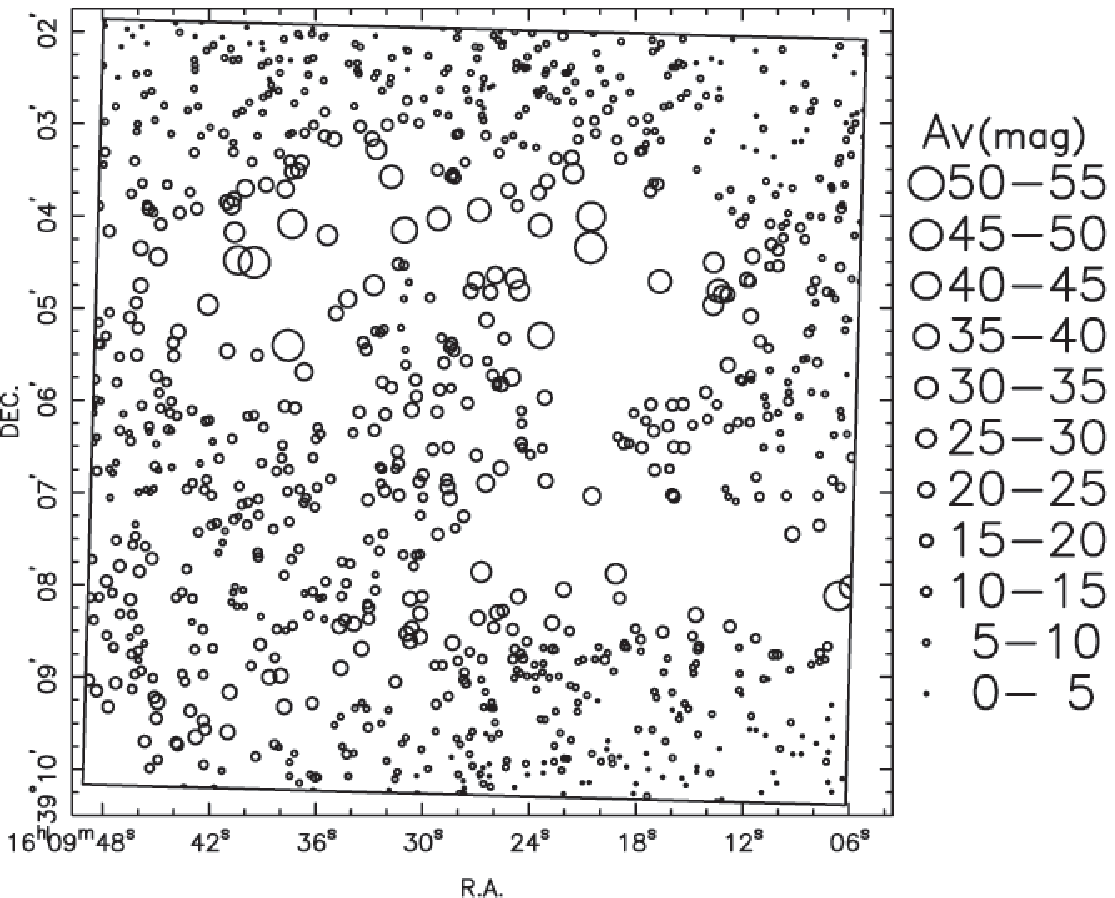}}
  \end{center}
  \caption{The $A_V$ toward the background stars. The radius of circles represent the $A_V$. 
   }\label{fig:fig2}
\end{figure}

\section{Results and Discussion}

\subsection{Surface brightness and column density}

\begin{figure}
  \begin{center}
  \scalebox{0.8}{\includegraphics{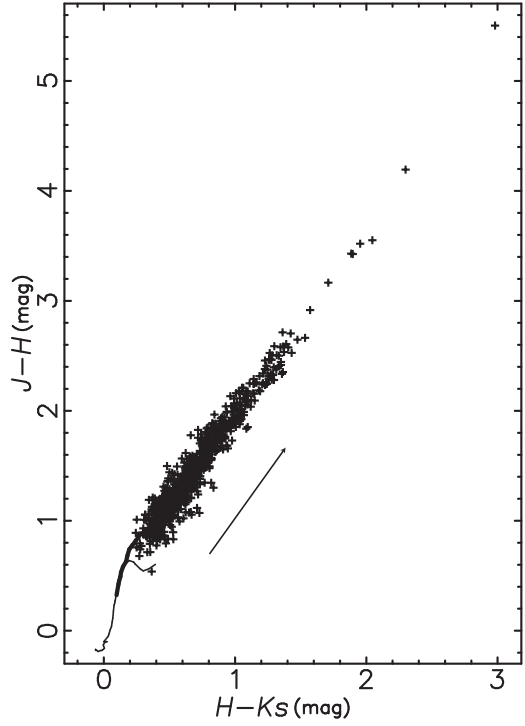}}
  \end{center}
  \caption{A $J-H$ vs. $H-K_s$ color-color diagram for the background stars.  
The thick and thin curves denote the loci of giants and dwarfs, respectively (Bessell \& Brett 1988). 
The arrow is parallel to a slope of $E(J-H)/E(H-K_s)$=1.7. The length of the arrow represents 
an amount of reddening with $A_V$=10 mag which is calculated from vdH. 
    }\label{fig:fig3}
\end{figure}

\begin{figure}
  \begin{center}
  \includegraphics{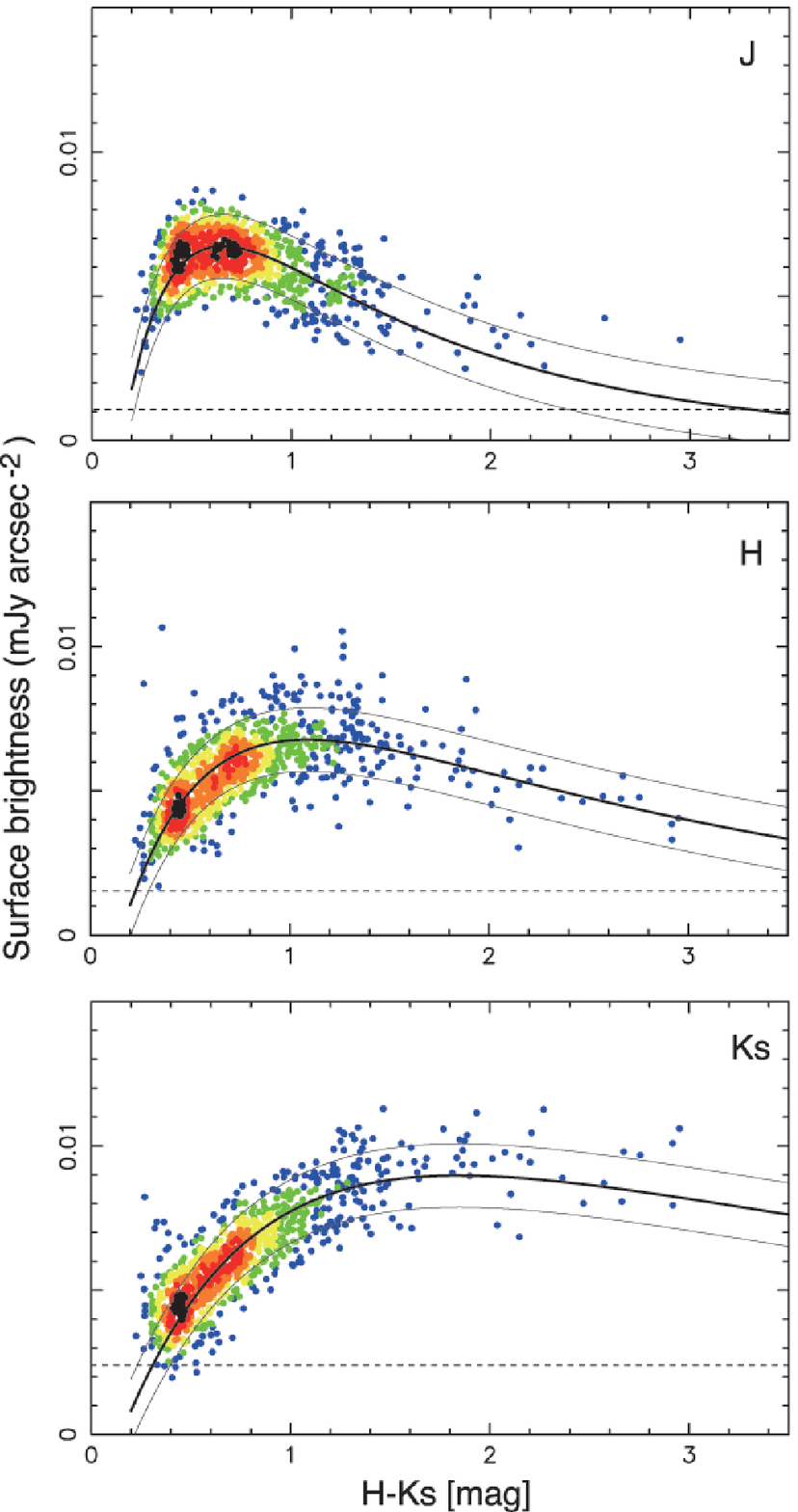}
  \end{center}
  \caption{Relationship between $H-K_s$ and surface brightness toward background stars at $J$, $H$, and $K_s$. 
  Colors of the points denote number density of the plot in the diagram; blue, green, yellow, orange, red, and black 
  from low to high. The 3-sigma limiting fluxes are indicated by dashed lines.
    }\label{fig:fig4}
\end{figure}

Surface brightness maps in the $J$, $H$, and $K_s$ bands are shown in Figure \ref{fig:fig1}. 
The values of $A_V$ toward the background stars are shown in Figure \ref{fig:fig2}. 
The $A_V$ is estimated from  the observed $H-K_s$ color of each background star 
by $A_V = 18.0 \times [(H-K_s) -  (H-K_s)_{\rm intrinsic}]$. The conversion factor between $A_V$ and the color excess 
is calculated from the theoretical curve No. 15 of  van de Hulst (1946, hereafter vdH).
\footnote{ 
Nishiyama et al. (2006) derived the ratios of total to selective extinction in the $J$, $H$, and $K_s$ bands 
toward the Galactic center with IRSF/SIRIUS. The study used a number of red clump stars toward 
the Galactic center region and the results are quite reliable. 
Relations among the $V$ band and NIR bands which are consistent with Nishiyama et al. (2006) are not yet obtained. 
The results of Nishiyama et al. (2006) are significantly different from the values of Rieke \& Lebofsky (1985) 
but are close to ones calculated from the curve No.15 of vdH. 
Thus, we adopted the curve No. 15 of vdH for calculations of factors for the conversions and extinction ratio in this paper. 
}
Here, we set $(H-K_s)_{\rm intrinsic} = 0.15$ for all the background stars, which is the average 
value of those behind the Lupus 3 dark cloud as estimated in Nakajima et al. (2003).  
Figure \ref{fig:fig3} shows a color-color diagram for the background stars whose fluxes were well determined 
in all the three bands. While the Lupus 3 dark cloud is a low-mass star forming region, 
the observed field contains no stars which have a clear intrinsic color excess; colors of all the stars can be 
explained by reddening of dwarfs or giants. 
We therefore assume that the color excess of the majority of the stars are totally due to extinction by dust, 
although stars without the $J$ magnitude have not been examined by the color-color diagram. 
We compared the observed $H-K_s$ color and the surface brightness in each band toward each background star. 
We show the results in Figure \ref{fig:fig4}. We used only stars with the color error of $\sigma_{H-K_s} < 0.14$ mag and 
surface brightness (i) with flux error less than 0.002 mJy arcsec$^{-2}$ and 
(ii) with flux larger than the 3-sigma limiting flux at each band. 

There is a well defined relation between the observed $H-K_s$ color and the surface brightness for each band. 
A simple interpretation of the relation is as follows. 
Forward scattering is dominant and we consider that the starlight behind the cloud is the main source of 
the surface brightness in this case. 
Surface brightness has a peak where the column density has an adequate value so that there is enough amount of dust 
to scatter the incident background star light into diffuse light, while the scattered light suffers from only moderate extinction. 
Surface brightness is faint where the column density is small, because incident light is less scattered and 
therefore is less turned into diffuse light. 
Surface brightness is faint again where the column density is large, because incident light and scattered light suffer from heavy extinction. 
The value of $H-K_s$ for the peak surface brightness depends on the wavelength; at longer wavelengths the peak surface 
brightness takes place at higher $H-Ks$.

The surface brightness is, technically, a result of complex processes among dust grains and incident starlight, 
which depend on optical characteristics of dust and cloud geometry. 
In the followings, however, we do not take such complex physical processes into account;  
we consider the dark cloud as a {\it black box} and 
we try to obtain an empirical relationship between the observed $H-K_s$ color and 
the surface brightness for each band quantitatively and to estimate 
the column density based on the surface brightness toward any direction even without background stars. 

We consider a simple one-dimensional scattering model to make an adequate fitting function to obtain an empirical 
relationship. We assume that a dark cloud is illuminated uniformly from the behind of the cloud and 
forward scattering is dominant at the wavelengths. 
When a beam of intensity $I_\lambda$ passes through scattering and absorbing material of thickness $ds$, 
the radiation transfer equation for isotropic scattering is generally given by 
\begin{equation}
dI_{\lambda} / ds = - (\alpha_{\lambda} + \sigma_{\lambda}) I_{\lambda} + \sigma_{\lambda} J_{\lambda} . 
\end{equation}
Here, $\alpha_{\lambda}$ is the absorption coefficient, $\sigma_{\lambda}$ is the scattering coefficient, 
and $J_{\lambda}$ is the mean intensity within the scattering material. 
We replace $J_{\lambda}$ with $g_\lambda ' I_\lambda$ 
based on the assumptions of one-dimensional symmetry and strong forward scattering. 
Here $g_\lambda '$ is the probability of radiation being scattered in the forward directions. 
If we define  
\begin{equation}
I(\Theta)_{scat} = \int^\Theta_0 d\theta \int^{2\pi}_0 d\phi \; I_{scat} (\theta, \phi) \sin\theta \cos\phi, 
\end{equation}
the whole radiation scattered in the forward directions, $I(\pi/2)_{scat}$, contributes to the source term due to scattering. 
Here, $\theta$ and $\phi$ are cylindrical coordinates together with the axis parallel to $I_\lambda$. 
The coefficient $g_\lambda '$ equals to $I(\pi/2)_{scat}/I(\pi)_{scat}$. 
Then we get 
\begin{equation}
dI_{\lambda} / ds = - \kappa_\lambda I_\lambda + g_\lambda '  \gamma_\lambda \kappa_\lambda I_\lambda . 
\end{equation}
Here, $\kappa_\lambda = \alpha_{\lambda} + \sigma_{\lambda}$ is the total opacity 
and $\gamma_\lambda = \sigma_{\lambda} / (\alpha_{\lambda} + \sigma_{\lambda})$ is the  albedo. 

The observed intensity of scattered radiation is then 
\begin{equation}
I_\lambda = \int_{I_{\lambda0}} ^{I_\lambda} dI_{\lambda} '  - I_{\lambda0}  exp(-\tau_\lambda) 
 = I_{\lambda0} [exp(g_\lambda '  \gamma_\lambda \tau_\lambda) - 1] exp(-\tau_\lambda) . 
\end{equation}
Here, $I_{\lambda0}$ is the average surface brightness of initial intensity of radiation of background stars and   
$\tau_\lambda =  \int \kappa_\lambda ds$ is optical depth along the line of sight. 
Using the relationships $A_J = 4.39 \ E(H-K_s)$, $A_H = 2.55 \ E(H-K_s)$, and $A_{K_s} = 1.58 \ E(H-K_s)$ 
from the extinction curve No. 15 of vdH, 
we write $\tau_\lambda = 0.921 A_\lambda$ using $H-K_s$. 
Consequently, we get the following equation with two free parameters, $I_{\lambda0}$ and $f_d$. 
\begin{equation}
I_\lambda = I_{\lambda0} \ \{ exp [f_d \times \alpha_\lambda (H-K_s - 0.15)] -1 \} \ exp[ -1 \times \alpha_\lambda (H-K_s - 0.15)]
\end{equation}
Here, $\alpha_\lambda$ is $\tau_\lambda /  (H-K_s - 0.15)$ and equals to 4.04, 2.35 and 1.46 for $J$, $H$, and $K_s$, respectively. 
We set $(H-K_s)_{\rm intrinsic} = 0.15$ for this equation as well as noted above. 
The parameter $f_d$ represents the product of $g_\lambda '$ and  $\gamma_\lambda$. 

We obtained plausible fittings to the $H-K_s$ and surface brightness relationship with the equation (5).
The fitting parameters were summarized in Table \ref{tab:first}. 
The thick curve denotes the best fit for each band in Figure \ref{fig:fig4}. 
We calculated standard deviation of the surface brightness  
from the best fit curve for each band and obtained 0.001 mJy arcsec$^{-2}$ for all the band. 
The upper and lower thin curves denote the best fit curve $\pm$ 0.001 mJy arcsec$^{-2}$. 
We consider that the area between the two thin curves defines the empirical relationship 
between the surface brightness and $H-K_s$, i.e., dust column density, in the diagram for each band. 

\begin{table}
  \caption{Fitting parameters.}\label{tab:first}
  \begin{center}
    \begin{tabular}{ccc}
    \hline
      band & $I_{\lambda0}$  & $f_d$ \\
       &  (mJy arcsec$^{-2}$) &   \\ 
    \hline
      $J$ & 0.0123  & 0.808  \\
      $H$ & 0.0113 & 0.845  \\
      $K_s$ & 0.0133 & 0.889 \\    
     \hline
    \end{tabular}
  \end{center}
\end{table}

\subsection{Estimate of column density}

\begin{figure}
  \begin{center}
  \scalebox{0.8}{\includegraphics{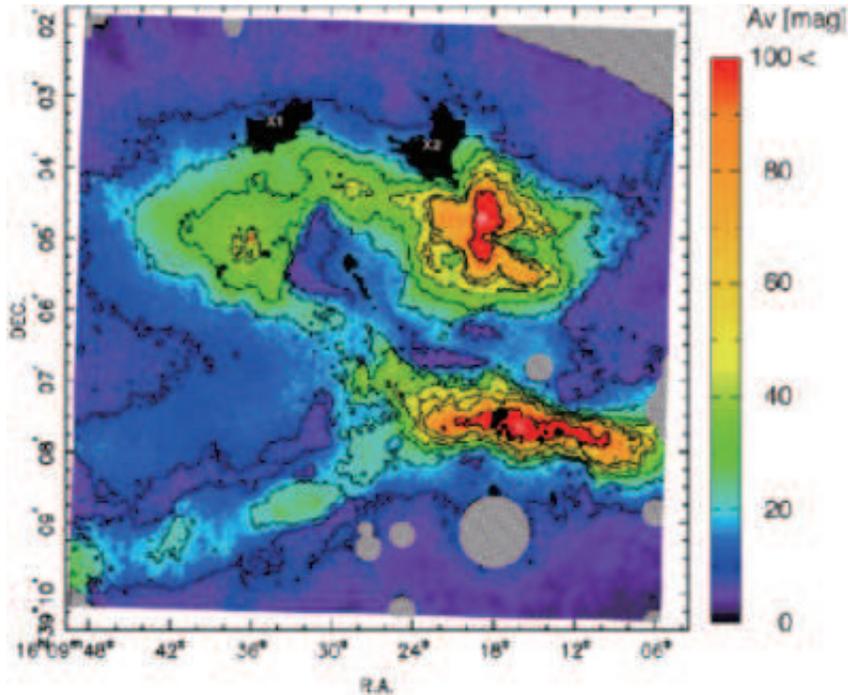}}
  \end{center}
  \caption{$A_V$ map estimated from the surface brightnesses. Color scale indicates $A_V$ in magnitude. 
  The contour levels are 10, 20, 30, 40, 50, 60 and 100 mag. The pixels suffering from bad pixel clusters and saturated 
  stars are hatched. There are three dense cores A, B, and C. The nomenclature is the same as in Nakajima et al. (2003). 
  }\label{fig:fig5}
\end{figure}

We estimated a column density for each pixel from the surface brightnesses  
by using the equation (5) and using the parameters in Table 1.  
The resultant $A_V$ map is shown in Figure \ref{fig:fig5}. The procedure is as follows. 
Suppose that a pixel has surface brightnesses of $F_J \pm \sigma_{F_J}$, $F_H \pm \sigma_{F_H}$, 
and $F_{K_s} \pm \sigma_{F_{K_s}}$ at $J$, $H$, and $K_s$, respectively. 
For example, in the $J$ band graph of Figure \ref{fig:fig4}, intersections of 
$y=F_J$ and the empirical relationship give solutions of $x=H-K_s$ for the $J$ band. 
Their errors are given by intersections of $y=F_J \pm \sigma_{F_J}$ and {\it the curved empirical relationship band}. 
When $y=F_J$ and the empirical relationship do not intersect but 
$y=F_J \pm \sigma_{F_J}$ and {\it the curved empirical relationship band} do intersect, 
a solution is given as the most likely point defined by the two areas;  
we assumed that each function has a gaussian probability distribution along the $y$-direction 
with $\sigma$=$\sigma_{F_J}$ or the standard deviation of 0.001 mJy arcsec$^{-2}$, and calculated 
the product of the two gaussians to find the point with the maximum probability in the x-y plane. 
We obtained solutions of $H-K_s$ for the pixel by using the surface brightnesses at $H$ and $K_s$ in the same method. 
A surface brightness can correspond to two $H-K_s$ ranges for a band, i.e., there can be two solutions, 
because the relationship is not monotonic but has a peak in the range considered. 
We checked all the possible combinations of the solutions for the $J$, $H$, and $K_s$ bands 
and if the solutions for a combination are consistent with each other within their errors, we considered them 
as the right combination and we calculated the average of the solution and the propagated error 
to obtain the final solution and error of $H-K_s$ for the pixel. 
If the surface brightnesses for one or two bands are less than detection limit, we calculated the solutions and errors 
only from the other bands. Such cases occurred at areas with large extinction marked as B and C in Figure \ref{fig:fig5} 
and at areas with small extinction. In such cases, from the two solutions we selected the larger one for areas with 
large extinction and the smaller one for areas with small extinction. 
Consequently, we converted the $H-K_s$ to $A_V$ by using $A_V = 18.0 \times [(H-K_s) -  (H-K_s)_{\rm intrinsic}]$ 
in which we set $(H-K_s)_{\rm intrinsic} = 0.15$ as we noted above.
We calculated a median of 5 $\times$ 5 pixels around each of pixels 
and set the median as the solution for the pixel. 
There are some clusters of pixels which have no solutions. 
We indicate the pixels without solutions in black in Figure \ref{fig:fig5}. 
A pixel with a larger $A_V$ tends to have a larger error. 
We calculated the mode of the error for every 5 mag bin along $A_V$. 
The median of the errors at some $A_V$ are given in Table 2. The error is typically $\sim$ 30\%. 

There are two outstanding black areas at the northern edge of the dark cloud, which are indicated as X1 and X2 
in Figure \ref{fig:fig5}. They have no solutions of $A_V$, which means that the surface brightness cannot be explained 
by the illumination of background stars. It is possible that there are other nearby illuminating sources contributing 
to the flux of the two areas. Figure \ref{fig:fig1} shows that the two areas have the brightest flux especially in the $H$ band.  
The flux of area X2 is likely attributed to the young stellar object embedded in the core B. 
The young stellar object is associated with HH78 and a probable jet (Nakajima et al. 2003, Teixeira, Lada, \& Alves 2005). 
As noted in Nakajima et al. (2003), there are two possible cavity structures created by the young stellar jet. 
One is toward west and appears as a fan shaped local minima of $A_V$ $\sim$ 30 mag 
in Figure \ref{fig:fig5}. The other corresponds to the area X2. The area X2 might be illuminated by the young 
stellar object. The illuminating source for the area X1 is not clear. 

The fitting of the equation (5) is obtained from the data with $H-K_s < 3$, which means that the conversion from 
the surface brightnesses into $A_V$ in the range of $A_V$ $>$ 50 mag is based on extrapolation. 
Therefore, the relationship is, technically, applicable only for $A_V$ $<$ 50 mag. 
The empirical relations tend to underestimate the $J$- and $K_s$-band surface brightnesses at high extinctions, 
which may also affect the reliability in the estimate of $A_V$ at high extinctions. 
In order to evaluate accuracy of the fitting at high extinctions, we tried another procedure with which the fitting  
is more optimized at high extinctions for the $J$ and $K_s$ bands. 
We separated the data into two groups: those with $H-K_s \le 1$ and those with $H-K_s > 1$.  
We searched the fitting parameters which minimize the sum of the reduced chi-squares for the two groups. 
We obtained the best parameters $(I_{\lambda0}, f_d)$=(0.01022, 0.8083) and (0.0098, 0.9654) for $J$ and $K_s$, respectively. 
Constant of 0.001 mJy arcsec$^{-2}$ was needed to add the equation (5) for the best fittings for both $J$ and $K_s$. 
The resultant relations do not underestimate the surface brightnesses at $J$ and $K_s$.  
We calculated the difference of $A_V$ between this fitting procedure and the former one for each pixel. 
In a range of $0 \le A_V \le 50$, the fraction of pixels which have the difference of $A_V$ less than $A_V$ error 
is 96\%.  The fraction is 54\% and 16\% in ranges of $50 \le A_V \le 100$ and $100 \le A_V \le 150$, respectively. 
Thus, the estimated $A_V$ and error are marginally unreliable in ranges of $50 \le A_V \le 100$ and 
completely unreliable for $100 \le A_V \le 150$.

\begin{figure}
  \begin{center}
  \scalebox{0.8}{\includegraphics{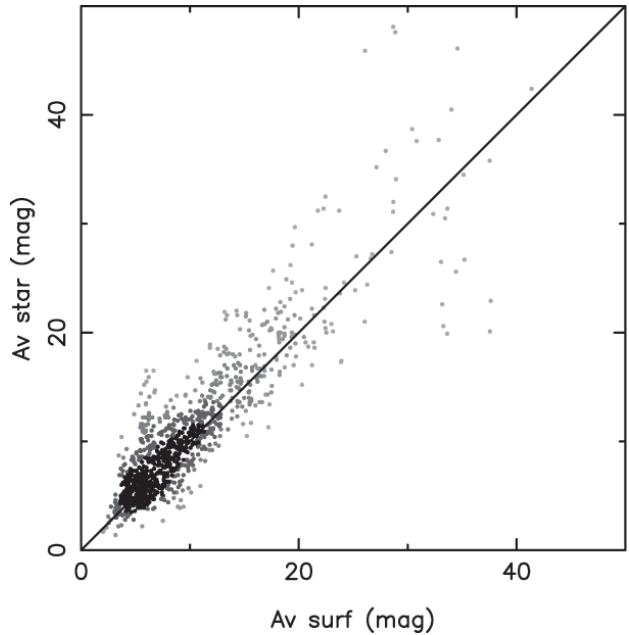}}
  \end{center}
  \caption{Relationship between the $A_V$ estimated from the surface brightnesses and that 
from $H-K_s$ of background stars. 
The grayscale of the points are assigned according to the number density on the graph; darker for higher density. 
The solid line has a slope of 1.
    }\label{fig:fig6}
\end{figure}

Figure \ref{fig:fig6} shows the relationship between the $A_V$ estimated from the surface brightnesses and that 
from $H-K_s$ of background stars, in other words, a comparison of Figures 2 and 5. 
There is a slight tendency that $A_V$ estimated from the background starts is larger than that from the surface 
brightnesses. A least square fit yields a slope of 1.05 for the relationship. 
This may be due to the shadowing effects produced by optically thick regions (Juvela et al. 2006).
However, the deviation of 5 \% is not significant compared to the typical error of 30 \% for the $A_V$ estimate 
with the surface brightnesses.

\begin{table}
  \caption{The median of error.}\label{tab:second}
  \begin{center}
    \begin{tabular}{cc}
    \hline
      $A_V$ (mag) & median of error (mag)\\
    \hline
10 & 3.5 \\
20 & 6.4 \\
30 & 8.0 \\
40 & 11 \\
50 & 14 \\
60 & 17 \\
70 & 21 \\
80 & 25 \\
90 & 28 \\
100 & 33 \\
     \hline
    \end{tabular}
  \end{center}
\end{table}

\subsection{Fitting parameters}

We made a simple one-dimensional formula with two parameters. 
In reality, however, the observed cloud is not one-dimensional, and  
the scattered source term in the equation (3) is to be altered according to the local cloud density 
structure and radiation field. In this study, the derivation of column densities is essentially empirical.  
The fitted values of the parameters do not necessarily match the actual 
physical parameters: intensity of background radiation and dust properties.   
Here, we examine the fitted values.

We compared the fitting parameter $I_{\lambda0}$ and the average surface brightness from the background stars for each band. 
We made a $K_s$ band luminosity function of the point sources with $A_{Ks} \le 1.0$ mag, 
in which magnitude was de-reddened by $K_{s0}=K_s - 1.58 (H-K_s - 0.15)$. 
The log of the luminosity function is well fitted by a power-law in the magnitude range of $10 \le K_s \le 17$. 
We calculated a total flux from the de-reddened $K_s$ magnitude between $10 \le K_s \le 17$.  
In order to correct the total flux by taking the fainter stars into account, 
we also estimated the flux contributed by background stars with $17 \le K_s \le 30$ 
by integrating the power-law fitting function. 
Then we divided the total flux by the area of $H-K_s \le 0.78$ which is estimated by the $A_V$ map in Figure 5. 
The color of $H-K_s = 0.78$ corresponds to $A_{Ks} = 1.0$ mag.  
We obtained 0.011 mJy arcsec$^{-2}$, with an uncertainty of a factor of 2,  
as the average flux at $K_s$ from the background stars. 
Similarly, we estimated the average flux at $H$ to be 0.011 mJy arcsec$^{-2}$. 
Thus, the fitting parameter $I_{\lambda0}$ is consistent with the average flux from the background stars in the $H$ and $K_s$ band. 
We could not obtain a meaningful value of the average flux for the $J$ band, because the number of the stars 
with $A_J \le 1.0$ mag is too small and the luminosity function was not fitted by a power-law function.  

The parameter $f_d$ is a product of $\gamma_\lambda$ and $g_\lambda '$: 
the albedo and the probability of radiation being scattered into the forward direction. 
Both $\gamma_\lambda$ and $g_\lambda '$ range between 0 and 1. 
The resultant values of $f_d=$0.81-0.89 means that both the parameters 
have values close to the upper limit. 
This is consistent with large albedo obtained in Nakajima et al. (2003).  

\subsection{Comparison with the previous studies}

Padoan et al. (2006) and Juvela et al. (2006, 2008) developed a new method of mapping 
column density of dark cloud using NIR scattered light. 
This paper is based on the similar considerations; however, this paper has some new aspects.   
Our method uses an empirical formula whose parameters are directly calibrated 
by the color excess of the background stars, while the previous studies used an analytic formula 
and used an extinction map derived from the color excess just for comparison as an independent tracer. 
We do not use a smoothed extinction map but the color excess of individual stars for comparison 
with the surface brightness toward each star. 
The use of individual background stars avoids the problems with the bias when there are 
few background stars (Juvela et al. 2008). 
By adopting a different formula, our method covers a part beyond the point where surface 
brightness starts to decrease with increasing column density, while the previous studies did 
for only the non-saturated part of the relation.

\subsection{Other remarks}

When this method is applied to other clouds, the fitting parameters should be determined for each cloud. 
The fitting parameters for each band in the equation (5) can be different among dark clouds. 
The parameter $I_0$ represents the average surface brightness of background stars, which 
depends on the Galactic coordinates. The parameter $f_d$ represents optical property of dust grain 
in the cloud, which can be different among the clouds. 

Teixeira, Lada, \& Alves (2005) made an extinction map toward the Lupus 3 dark cloud with 
the $H-K_s$ color of the background stars. While their map generally agrees with ours (see their Figure 2 and 6) 
in their smoothing scale, 30 and 40 arcsec, the map failed to trace the core C 
(the core E in their paper) properly due to the lack of stars.

\section{Conclusion}

We constructed an empirical relationship equation between the surface brightness 
and column density of the Lupus 3 dark cloud for each $J$, $H$, and $K_s$ band. 
By using the equations, we obtained a column density map with a pixel scale of 2.3 $\times$ 2.3 
arcsec$^2$ and a large dynamic range up to $A_V$ = 50 mag of the cloud from the NIR surface brightness. 
We expect the empirical method to be one of the new tools of column density mapping of dark clouds. 

\section{Acknowledgment }

We thank the anonymous referee for the constructive comments. 
We thank Dr. Daisuke Kato for his comments. 
RK and MT are supported by Grants-in-Aid from the Ministry of Education, Culture, Sports, 
Science and Technology (No 16340061).  
This publication makes use of data products from the Two Micron All Sky Survey, 
which is a joint project of the University of Massachusetts and the Infrared Processing 
and Analysis Center/California Institute of Technology, funded by the National Aeronautics 
and Space Administration and the National Science Foundation.

\end{document}